\documentstyle[12pt,a41,epsfig,cite,amssymb,amstex]{article}

 \newcommand{\N}{\nonumber}

\newcommand{\bea}{\begin{eqnarray}}
\newcommand{\bq}{\begin{equation}}
\newcommand{\eea}{\end{eqnarray}}
\newcommand{\eq}{\end{equation}}

\newcommand{\gsim}{\raisebox{-0.07cm   }
{$\, \stackrel{>}{{\scriptstyle\sim}}\, $}}

\newcommand\be{\begin{eqnarray}}
\newcommand\ee{\end{eqnarray}}

\newcommand\ep{\varepsilon}

\begin{document}
\noindent
\sloppy
\thispagestyle{empty}

\begin{flushleft}
DESY 08-187 
\\
SFB/CPP--08--107\\
IFIC/08--68\\
December 2008
\end{flushleft}

\vspace*{\fill}
\begin{center}
{\LARGE\bf  The Gluonic Operator Matrix Elements} 

\vspace{2mm}
{\LARGE\bf at \boldmath{$O(\alpha_s^2$)} for DIS Heavy Flavor Production}

\vspace{2cm}
\large
Isabella Bierenbaum~\footnote{Present address: Instituto de Fisica
Corpuscular, CSIC-Universitat de Val\`{e}ncia, Apartado de Correros 22085,
E-46071 Valencia, Spain.},
Johannes Bl\"umlein  and Sebastian Klein
\\
\vspace{2em}
\normalsize
{\it Deutsches Elektronen--Synchrotron, DESY,\\
Platanenallee 6, D--15738 Zeuthen, Germany}
\\
\vspace{2em}
\end{center}
\vspace*{\fill}
%
\begin{abstract}
\noindent
We calculate the $O(\alpha_s^2)$ gluonic operator matrix elements for the 
twist--2 operators, which contribute to the heavy flavor Wilson coefficients 
in unpolarized deeply inelastic scattering in the region $Q^2 \gg m^2$, up 
to the linear terms in the dimensional parameter $\varepsilon$, ($D= 
4 + \varepsilon$).  These quantities are required for the description of parton 
distribution functions in the variable flavor number scheme (VFNS). The 
$O(\alpha_s^2 \varepsilon)$ terms contribute at the level of the $O(\alpha_s^3)$ 
corrections through renormalization. We also comment on additional terms, which 
have to be considered in the fixed (FFNV) and variable flavor number scheme, 
adopting the $\overline{\rm MS}$ scheme for the running coupling constant.
\end{abstract}
\vspace*{\fill}
\newpage
\section{Introduction}
\label{sec-1}

\vspace{1mm}\noindent
Both unpolarized and polarized deep-inelastic structure functions receive 
contributions from light partons and heavy quarks. In the unpolarized 
case, the charm 
quark contribution may amount to 25-35\% in the small $x$ region,~\cite{JBSR}.
Since the scaling violations in case of the heavy quark contributions differ
significantly from those of the light partons in a rather wide range starting 
from lower values of $Q^2$, a detailed description of the heavy quark 
contributions is required. In the FFNS the corresponding Wilson coefficients
were calculated to next-to-leading order (NLO) in a semi-analytic approach in 
\cite{NLO}.~\footnote{
For a fast implementation of these corrections in Mellin space see \cite{SAJB}.}
Consistent QCD analyzes to 3-loop order require the description of both the
light and the heavy flavor contributions at this level to allow for an 
accurate measurement of the QCD scale $\Lambda_{\rm QCD}$ in singlet 
analyzes~\cite{HERA}~\footnote{For a determination of $\Lambda_{\rm QCD}$ 
effectively analyzing the scaling violations of the non-singlet world data to 
$O(a_s^4)$, ($a_s = \alpha_s/(4\pi)$), 
cf. Ref.~\cite{BGR}.} and the measurement of the parton distribution 
functions. The calculation of the 3-loop heavy flavor Wilson coefficients in 
the whole $Q^2$ region is currently not within reach. However, as noticed in 
\cite{BUZA1}, a very precise description of the heavy flavor Wilson coefficients 
contributing to the structure function $F_2(x,Q^2)$ is obtained for $Q^2 \gsim 
10~m^2_Q$, disregarding the power corrections $\propto (m_Q^2/Q^2)^k, k \geq 1$, 
which covers the main region for 
deep--inelastic physics at HERA. In this case, the Wilson coefficients are even 
obtained in analytic form. The heavy flavor 
Wilson coefficients factorize into universal massive operator matrix 
elements (OMEs) $A_{ij}(\mu^2/m^2_Q)$ and the light flavor Wilson coefficients 
$c_j(Q^2/\mu^2)$ \cite{WILS} in this limit,
\begin{eqnarray}
H_i(Q^2/\mu^2,m^2_Q/\mu^2,z) = A_{ji}(m^2_Q/\mu^2,z) \otimes c^j(Q^2/\mu^2,z),
~~~~i = 2,L~.
\end{eqnarray}
Here, $\mu$ denotes the factorization scale and $z$ the longitudinal momentum 
fraction of the parton in the hadron. 

In the strict sense, only massless particles can be interpreted as partons in 
hard scattering processes since the lifetime of these quantum-fluctuations off
the hadronic background $\tau_{\rm life} \propto 1/(k_\perp^2+m^2_Q)$ has to be 
large 
against the interaction time $\tau_{\rm int} \propto 1/Q^2$ in the infinite 
momentum frame, \cite{DY}. In the massive case, $\tau_{\rm life}$ is 
necessarily finite and there exists a larger scale $Q^2_0$ below which any 
partonic 
description fails. From this it follows, that
the heavy quark effects are genuinely described by the {\sf process dependent}
Wilson coefficients. Since parton-densities are {\sf process independent}
quantities, only those pieces out of the Wilson coefficients can be used
to define them for heavy quarks at all. Clearly this is impossible in 
the region close to threshold but requires $Q^2/m^2_Q = r \gg 1$, with $ r 
\gsim 
10$ in case of $F_2(x,Q^2)$. For $F_L(x,Q^2)$ the corresponding ratio 
even turns out to be $r \gsim 800$,~\cite{BUZA1,JBFL,GRS}. Heavy flavor parton 
distributions can thus be constructed only for scales $\mu^2 \gg m^2_Q$. Their
use in observables is restricted to a region, in which the power 
corrections can be safely neglected. This range may strongly depend 
on the observable considered as the  examples of $F_2$ and $F_L$ show.

For processes in the high $p_\perp$ region at the LHC, in which the above 
conditions are fulfilled,  one may use heavy flavor parton distributions
by proceeding
as follows. In the region $Q^2 \gsim 10 m^2_Q$ the heavy flavor contributions 
to the $F_2(x,Q^2)$--world data are very well described by the asymptotic 
representation in the FFNS. For large scales one can then form a variable 
flavor representation including one heavy flavor distribution, \cite{BUZA2}.
This process can be iterated towards the next heavier flavor, provided the 
{\sf universal} representation holds  and all power corrections can be safely 
neglected. One has to take special care of the fact, that
the matching scale in the coupling constant, at which the 
transition $N_f \rightarrow N_f+1$ is to be performed, often differs rather
significantly from $m_Q$, cf. \cite{BN1},

For the procedure outlined above, besides the quarkonic heavy
flavor OMEs \cite{BUZA1,BBK1},
the gluonic matrix elements are required. These have been calculated 
to $O(a_s^2)$ in Ref.~\cite{BUZA2}. 
Here we verify this calculation and extend it to the terms of 
$O(a_s^2 \varepsilon)$, which enter the $O(a_s^3)$ matrix elements through 
renormalization. The corresponding contributions for the quarkonic matrix 
elements were calculated in \cite{BBK2}.~\footnote{For the first few values 
of the 
Mellin moment $N$ the pure-singlet and non-singlet quarkonic OMEs were 
calculated to $O(a_s^3)$ in Refs.~\cite{THRL}.}  

The paper is organized as follows. In Section~2 we summarize the relations
needed to describe heavy flavor parton densities out of parton densities
of only light flavors in terms of massive operator matrix elements. 
Furthermore, we point out  terms to be added to the FFNS description in
the $\overline{\rm MS}$ scheme if compared to \cite{NLO,BUZA1}, which are of 
numerical 
relevance, cf. \cite{ABK}. We also comment on the question of the effective 
number of flavors
considering the renormalization of the process. In Section~3 the 
massive gluonic 2--loop operator matrix elements are presented and Section~4
contains the conclusions.
\section{Heavy Flavor Parton Densities}
\label{sec-2}

\vspace{1mm}\noindent
In the asymptotic region $Q^2 \gg m_Q^2$ one may define heavy flavor parton 
densities. This is done under the further assumption that for the other
heavy flavors the masses $m_{Q_i}$ form a hierarchy $m_{Q_1}^2 \ll m_{Q_2}^2 
\ll~~{\rm etc.}$ Allowing for {\sf one} heavy quark of mass $m_Q$ and $N_f$ 
light quarks one obtains the following light and heavy-quark parton 
distribution functions in Mellin space, 
\cite{BUZA2},
\begin{eqnarray}
\label{HPDF1}
f_k(N_f+1,\mu^2,N) + f_{\overline{k}}(N_f+1,\mu^2,N)
&=& A_{qq,Q}^{\rm NS}\left(N_f,\frac{\mu^2}{m_Q^2},N\right)
\cdot \left[f_k(N_f,\mu^2,N) + f_{\overline{k}}(N_f,\mu^2,N)\right]
\nonumber\\ 
& & + \tilde{A}_{qq,Q}^{\rm 
PS}\left(N_f,\frac{\mu^2}{m_Q^2},N\right)
\cdot \Sigma(N_f,\mu^2,N)
\nonumber\\ 
& & + \tilde{A}_{qg,Q}^{\rm 
S}\left(N_f,\frac{\mu^2}{m_Q^2},N\right)
\cdot G(N_f,\mu^2,N),
\\
\label{fQQB}
f_Q(N_f+1,\mu^2,N) + f_{\overline{Q}}(N_f+1,\mu^2,N)
&=&
{A}_{Qq}^{\rm PS}\left(N_f,\frac{\mu^2}{m_Q^2},N\right)
\cdot \Sigma(N_f,\mu^2,N)
\nonumber\\ && \hspace*{-4mm}
+ {A}_{Qg}^{\rm S}\left(N_f,\frac{\mu^2}{m_Q^2},N\right)
\cdot G(N_f,\mu^2,N)~.
\end{eqnarray}
Here $f_k (f_{\bar{k}})$ denote the light quark and anti--quark densities, 
$f_Q 
(f_{\bar{Q}})$ the heavy quark densities, and 
$G$ the gluon density.
The flavor singlet, non-singlet and gluon densities for $(N_f+1)$ flavors are 
given by
\begin{eqnarray}
\Sigma(N_f+1,\mu^2,N) 
&=& \Biggl[A_{qq,Q}^{\rm NS}\left(N_f, \frac{\mu^2}{m^2_Q},N\right) +
          N_f \tilde{A}_{qq,Q}^{\rm PS}\left(N_f, \frac{\mu^2}{m^2_Q},N\right)
  \nonumber \\ &&
         + {A}_{Qq}^{\rm PS}\left(N_f, \frac{\mu^2}{m^2_Q},N\right)
        \Biggr]
\cdot \Sigma(N_f,\mu^2,N) \nonumber
\end{eqnarray} \begin{eqnarray}
& & \hspace*{-3mm} + \left[N_f \tilde{A}_{qg,Q}^{\rm S}\left(N_f, 
\frac{\mu^2}{m^2_Q},N\right) +
          {A}_{Qg}^{\rm S}\left(N_f, \frac{\mu^2}{m^2_Q},N\right) 
\right]
\cdot G(N_f,\mu^2,N) 
\\
\Delta(N_f+1,\mu^2,N) &=& 
  f_k(N_f+1,\mu^2,N)
+ f_{\overline{k}}(N_f+1,\mu^2,N) - \frac{1}{N_f+1} \Sigma(N_f+1,\mu^2,N)\\
\label{HPDF2}
G(N_f+1,\mu^2,N) &=& A_{gq,Q}^{\rm S}\left(N_f, \frac{\mu^2}{m^2_Q},N\right) 
                    \cdot \Sigma(N_f,\mu^2,N)
\nonumber\\
&& 
+ A_{gg,Q}^{\rm S}\left(N_f, \frac{\mu^2}{m^2_Q},N\right) 
                    \cdot G(N_f,\mu^2,N)~.
\end{eqnarray}
Here,
\begin{eqnarray}
\label{eqAij}
A_{ij}^{\rm NS (PS, S)} = \langle j| O_i^{\rm NS (PS, S)}|j \rangle 
 = \delta_{ij} + \sum_{l=1}^\infty a_s^l A_{ij}^{(l), \rm NS (PS, S)}
\end{eqnarray}
are the operator matrix elements of the local twist--2 non-singlet 
(NS), pure singlet (PS) and singlet (S) operators $O_j^{\rm NS 
(PS, S)}$ between on--shell partonic states $|j\rangle,~~j = q, g$ and
\begin{eqnarray}
A_{ij} = N_f \tilde{A}_{ij}~.
\end{eqnarray}
Note that in the pure-singlet case the term $\delta_{ij}$ 
in (\ref{eqAij}) is absent.
The normalization of the quarkonic and gluonic operators obtained in the light 
cone expansion can be chosen arbitrarily. 
It is, however, convenient to chose 
the relative factor such, that the non-perturbative nucleon-state expectation 
values, 
$\Sigma(N_f,\mu^2,N)$ and $G(N_f,\mu^2,N)$, obey
\begin{eqnarray}
\Sigma(N_f,\mu^2,N=2)+G(N_f,\mu^2,N=2) = 1
\end{eqnarray}
due to 4-momentum conservation. As a consequence, the OMEs fulfill the 
relations
\begin{eqnarray}
A_{qq,Q}^{\rm NS}(N_f,N=2)
   +N_f \tilde{A}_{qq,Q}^{\rm PS}(N_f,N=2)
   + \tilde{A}_{Qq}^{\rm PS}(N_f,N=2)
   +A_{gq,Q}^{\rm S}(N_f,N=2) &=& 1~, \label{sumrule1}
\\
N_f\tilde{A}_{qg,Q}^{\rm S}(N_f,N=2)
   +\tilde{A}_{Qg}^{\rm S}(N_f,N=2)
   + A_{gg,Q}^{\rm S}(N_f,N=2) &=& 1~.  \label{sumrule2}
\end{eqnarray}
The above scenario can be easily followed up to 2-loop order. Also here 
diagrams contribute which carry two different heavy quark flavors. At this 
level, the heavy degree of freedom may be absorbed into the coupling constant
and thus being decoupled temporarily. Beginning with 3-loop order the situation 
becomes more 
involved since there are graphs in which two different heavy quark flavors 
occur in nested topologies, i.e. the corresponding diagrams depend on the 
ratio $\rho = m_c^2/m_b^2$ yielding power corrections in $\rho$. There is no
strong hierarchy between these two masses. The above picture, leading to
heavy flavor parton distributions whenever $Q^2 \gg m^2_Q$ will not hold 
anymore, since one cannot decide immediately in case of the two-flavor 
graphs, whether they belong to the $c$-- or the $b$--quark distribution. 
Hence, the partonic description can only be maintained within a certain approximation 
by {\sf assuming} $\rho \ll 1$.

At this point we would like to add some remarks directed to readers who are 
not acquainted with the details of the calculation of heavy flavor Wilson 
coefficients.
In Refs.~\cite{NLO,BUZA1} the calculation to $O(a_s^2)$ was performed 
in a scheme, in which the heavy quark insertion in the external gluon
legs\footnote{In the present case, these are the terms $\propto~T_F^2$.}
were absorbed into the strong coupling constant through the relation
\begin{eqnarray}
\hat{a}_s &=& a_s(\mu^2) \left[1 + a_s(\mu^2) \delta a_s(\mu^2,m^2_Q)\right] \\
\label{DALS}
\delta a_s(\mu^2,m^2_Q) &=& S_\varepsilon \left\{\frac{2 
\beta_0(N_f)}{\varepsilon}
+ \frac{2 \beta_{0,Q}}{\varepsilon} 
\left(\frac{m^2_Q}{\mu^2}\right)^{\varepsilon/2}
\left(1+ \frac{\zeta_2}{8} \varepsilon^2+\frac{\zeta_3}{24}\varepsilon^3
 \right)\right\}~,
\end{eqnarray}
for $N_f$ light and one heavy flavor. Here $\hat{a}_s$ denotes the bare 
coupling constant $\hat{a}_s = \hat{g}_s^2/(16 \pi^2)$, $\beta_0(N_f) = (11/3) 
C_A 
- (4/3) T_F N_f, \beta_{0,Q} = - (4/3) T_F$,  $C_A = 3, T_F = 1/2$ for 
$SU(3)_c$. The spherical factor 
$S_\varepsilon = \exp[(\varepsilon/2)(\gamma_E - \ln(4\pi)]$ is
set  to one in the $\overline{\rm MS}$ scheme. 
The bare coupling constant is thus given by
\begin{eqnarray}
\label{eqASBARE}
\hat{a}_s &=& a_s(\mu^2) \left[1 + a_s(\mu^2) \frac{2 
\beta_0(N_f+1)}{\varepsilon} \right] - a_s^2(\mu^2) \beta_{0,Q} 
\ln\left(\frac{\mu^2}{m^2_Q}\right)+O(\ep)~.
\end{eqnarray}

\noindent
If the above scheme is applied, cf. \cite{BUZA1}, the renormalized OMEs
do not contain terms $\propto~T_F^2$. However, to express $a_s$ in the 
$\overline{\rm MS}$ scheme, only the first term in Eq.~(\ref{eqASBARE}) has 
to be used, while the second remains as a prefactor of the $1$--loop 
contributions. \footnote{The running coupling constant 
including heavy flavors in the MOM--scheme was presented in \cite{CKS} to 
$O(a_s^3)$ recently, showing Applequist--Carrazone \cite{AC} decoupling of the
heavy quark contributions.}
Hence, as has been lined out in \cite{BUZA2} later, the latter term 
appears e.g. in $A_{Qg}^{(2)}$ in front of $A_{Qg}^{(1)}$ in the case $Q^2 
\gg m_Q^2$. This is encountered 
as well in the present paper for $A_{gg,Q}^{(2)}$. Additionally, one
has to do the same for the complete heavy flavor Wilson coefficients, 
leading to the extra terms
\begin{eqnarray}
\label{eqCORR}
a_s^2(\mu^2) \beta_{0,Q} \ln\left(\frac{\mu^2}{m_Q^2}\right) 
H_{F_i}^{(1)}\left(
\frac{m^2_Q}{\mu^2},z\right) 
\end{eqnarray}
in the scattering cross section in the $\overline{\rm MS}$ scheme.
Here,
\begin{eqnarray}
H_{F_2}^{(1)}\left(\frac{m^2_Q}{Q^2},z\right) &=& 
       8T_F  \left\{\beta
          \left[ -\frac{1}{2} + 4z(1-z) + \frac{m^2_Q}{Q^2} z(2 z-1)
          \right] \right. \nonumber\\
          \hspace{-2cm}
         &+& \left. \left[-\frac{1}{2} + z - z^2 + 2 \frac{m^2_Q}{Q^2} z(3z-1)
            - 4 \frac{m^4}{Q^4} z^2 \right] \ln\left(\frac{1-\beta}{1+\beta}\right)
           \right\}~,
\\
H_{F_L}^{(1)}\left(\frac{m^2_Q}{Q^2},z\right) &=&  16 T_F \left[z (1-z) \beta -
\frac{m^2_Q}{Q^2} z^2 \ln\left|\frac{1+\beta}{1-\beta}\right|\right]~,
\\
\beta &=& \sqrt{1 - \frac{4 m^2_Q z}{Q^2(1-z)}}~,
\end{eqnarray}
denote the leading order Wilson coefficients for massive quarks with the 
strong coupling constant taken out.
In the same manner the contributions
$\propto T_F^2$ in the non-1PI-contribution in $A_{Qg}$, 
Ref.~\cite{BUZA2}, have to be removed, to avoid double counting if the 
asymptotic representation
for the heavy flavor Wilson coefficients is referred to.
Since the light flavor Wilson coefficients are calculated in the 
$\overline{\rm MS}$ scheme, the {\sf same} scheme has to be used for the 
massive OMEs. It should also be thoroughly used for renormalization, as the 
case for light flavors, to derive consistent results in QCD analyzes of 
deep-inelastic scattering data. 

In Refs.~\cite{NLO,BUZA1,BUZA2} another contribution, which belongs to
the inclusive heavy flavor contributions to the structure functions
$F_{2,L}(x,Q^2)$, was not 
dealt with. To $O(a_s^2)$
these are heavy quark loop insertions on the initial state gluon line for the
1st order light flavor Wilson coefficient $c_{2(L),g}^{(1)}(x,Q^2)$.
The corresponding contribution is
\begin{eqnarray}
\label{eqCORR1}
a_s^2(\mu^2) \beta_{0,Q} \ln\left(\frac{\mu^2}{m_Q^2}\right) 
c_{F_i,g}^{(1)}\left(z\right)~, 
\end{eqnarray}
see also \cite{CSN}. We also note that virtual corrections to 
$A_{qq,Q}^{(2),\rm NS}$, resp. $H_{2,L}^{(2), \rm NS}$, and $A_{gg,Q}^{(2),\rm S}$
need to be accounted for. In the asymptotic case $Q^2 \gg m^2_Q$ they lead
to $+$-functions, which regularize the soft singularity, 
cf.~\cite{BUZA1,BUZA2,BBK1,BBK2}. Here we always considered only one heavy 
quark contribution.  

The above expressions are derived for the FFNS. Charge-- and 
mass--renormalization are performed multiplicatively for the observables.
As evident from Eq.~(\ref{DALS}), $a_s(\mu^2)$ has to be 
calculated for $N_f +1$ flavors upon passing the $N_f+1$st flavor threshold.
In the FFNS the structure functions contain separate contributions of 
the strictly light and heavy flavors. The 
corresponding expressions for the Wilson coefficient contain anomalous 
dimensions which partly depend on $N_f$. In the case of the heavy flavor 
contributions to $O(a_s^2)$, \cite{BUZA1,BUZA2,BBK1,BBK2}, 
(\ref{HPDF1}--\ref{HPDF2}),  no  
closed light  fermion 
lines contribute, however. The evolution of the three light flavors proceeds with 
$N_f =3$. Due to this, there is {\sf no} arbitrariness in the choice 
of $N_f$ as sometimes anticipated in the literature.

\section{The Gluonic Operator Matrix Elements }
\label{Sec-AgqQ2L}
The description of heavy quark parton densities, 
Eqs.~(\ref{HPDF1}--\ref{HPDF2}), requires the massive operator matrix elements
given by the partonic on-shell expectation values $\langle 
p|O^K|p\rangle,~~p=q,g$,
of the operators, cf.~\cite{GRW},
\begin{eqnarray}
O^{F,0}_{\mu_1, \ldots, \mu_n} &=& i^{n-1} {\bf S} [\overline{\psi} 
\gamma_{\mu_1} D_{\mu_2} \ldots D_{\mu_n} \psi] - {\rm trace~terms}~, \\
O^{V}_{\mu_1, \ldots, \mu_n} &=& 2 i^{n-2} {\bf S} {\rm Sp}[F_{\mu_1 \alpha}
D_{\mu_2} \ldots D_{\mu_{n-1}} F_{\mu_n}^\alpha] - {\rm trace~terms}~. 
\end{eqnarray}
Here $D_\mu = \partial_\mu - i g_s t_a A_{\mu}^a$ denotes the covariant
derivative, $t_a$ are the generators of $SU(3)_c$, $\psi$ the quark 
fields, $A_\mu^a$ the gluon fields, $F_{\mu\nu}$ the gluonic field strength 
tensors, 
Sp the color trace, and ${\bf S}$ the operator which symmetrizes the Lorentz 
indices. The corresponding quarkonic operator matrix elements were calculated 
in Refs.~\cite{BUZA1,BBK1} to $O(a_s^2)$ and $O(a_s^2 \varepsilon)$ in 
\cite{BBK2}, respectively.

The renormalized gluonic operator matrix elements 
$A_{gq,Q}$ and $A_{gg,Q}$ to $O(a_s^2)$ are given by
\begin{eqnarray}
   A_{gq,Q}&=& a_s^2\Biggl[
                     \hat{A}_{gq,Q}^{(2)}
                    +Z_{gq}^{-1,(2)}(N_f+1)
                    -Z_{gq}^{-1,(2)}(N_f)
\nonumber\\ &&
                    +\Bigl(
                           \hat{A}_{gg,Q}^{(1)}
                          +Z_{gg}^{-1,(1)}(N_f+1)
                          -Z_{gg}^{-1,(1)}(N_f)
                      \Bigr)\Gamma_{gq}^{-1,(1)}
                     \Biggr] + O(a_s^3),
                            \label{AgqQRen2}
\\
    A_{gg,Q}&=&  a_s \left[
                         \hat{A}_{gg,Q}^{(1)}
                       +Z^{-1,(1)}_{gg}(N_f+1)
                       -Z^{-1,(1)}_{gg}(N_f) \right]
                           \label{AggQ1Ren1}
\nonumber\\ 
&& a_s^2  \Biggl[
                         \hat{A}_{gg,Q}^{(2)}
                       +Z^{-1,(2)}_{gg}(N_f+1)
                       -Z^{-1,(2)}_{gg}(N_f)
                       +Z^{-1,(1)}_{gg}(N_f+1)\hat{A}_{gg,Q}^{(1)}
\nonumber\\ &&
                       +Z^{-1,(1)}_{gq}(N_f+1)\hat{A}_{Qg}^{(1)}
                       +\Bigl[ \hat{A}_{gg,Q}^{(1)}
                              +Z^{-1,(1)}_{gg}(N_f+1)
                              -Z^{-1,(1)}_{gg}(N_f)
                        \Bigr] 
                             \Gamma^{-1,(1)}_{gg}(N_f)
\nonumber\\  &&+ \delta a_s \hat{A}_{gg,Q}^{(1)} \Biggr] + O(a_s^3).
 \label{AggQ2Ren1} 
\end{eqnarray}
Here $\hat{A}_{ij}$ are the operator matrix elements after mass--renormalization 
has been carried out. The $Z$--factors $Z_{ij}(N_f)$ 
renormalize the ultraviolet singularities of the operators and 
$\Gamma_{ij}(N_f)$
remove the collinear singularities, cf.~\cite{BUZA1,BUZA2,BBK1,BBK2}. The terms
$Z_{gq(g)}^{-1}(N_f+1)$ are equal to
{\small
\begin{eqnarray}
Z_{gq}^{-1}(N_f+1) &=& 
a_s\left[-\frac{1}{\varepsilon} \gamma_{gq}^{(0)}\right] 
+ a_s^2 \left[\frac{1}{\varepsilon} \left(-\frac{1}{2} \gamma_{gq}^{(1)} - 
\gamma_{gq}^{(0)} \delta a_s\right) + \frac{1}{\varepsilon^2}
\left(\gamma_{gq}^{(0)} \beta_0 + \frac{1}{2} \gamma_{gq}^{(0)} 
\gamma_{qq}^{(0)} + \frac{1}{2} \gamma_{gq}^{(0)} 
\gamma_{gg}^{(0)}\right)\right] 
\nonumber\\ &&
+ 
O(a_s^3)  \label{ZIgq}
\end{eqnarray} \begin{eqnarray}
Z_{gg}^{-1}(N_f+1) &=& 
1+a_s\left[-\frac{1}{\varepsilon} \gamma_{gg}^{(0)}\right] 
+ a_s^2 \left[\frac{1}{\varepsilon} \left(-\frac{1}{2} \gamma_{gg}^{(1)} - 
\gamma_{gg}^{(0)} \delta a_s\right) + \frac{1}{\varepsilon^2}
\left(\gamma_{gg}^{(0)} \beta_0 + \frac{1}{2} \gamma_{qg}^{(0)} 
\gamma_{gq}^{(0)} + \frac{1}{2} {\gamma_{gg}^{(0)}}^2 
\right)\right] 
\nonumber\\ &&
+ 
O(a_s^3)~. \label{ZIgg}
\end{eqnarray}
}
\normalsize

\noindent
In Eqs. (\ref{ZIgq},\ref{ZIgg}), $\gamma_{ij}^{(l)}$ are the $O(a_s^{l+1})$ 
anomalous dimensions and have to be taken -~as well as $\beta_0$~-
at $N_f+1$ flavors. We adopt the notation 
$\hat{\gamma}_{ij}^{(l)}= \gamma_{ij}^{(l)}(N_f+1)-\gamma_{ij}^{(l)}(N_f)$
and define for later use
\begin{eqnarray}
f(\varepsilon) = \left(\frac{m^2_Q}{\mu^2}\right)^{\varepsilon/2} \exp\left[
\sum_{k=2}^\infty \frac{\zeta_k}{k} 
\left(\frac{\varepsilon}{2}\right)^k\right]~. 
\end{eqnarray}
To the operator matrix element  $\hat{A}_{gg,Q}^{(1)}$ necessarily only 
non-1PI diagrams contribute. 
The un-renormalized OME $\hat{A}_{gq,Q}^{(2)}$ is given by~\footnote{In 
the following we drop the overall factor $[1+(-1)^N]/2$ in the operator matrix 
elements.}
  \begin{eqnarray}
   \hat{A}_{gq,Q}^{(2)}&=&\Bigl(\frac{m^2_Q}{\mu^2}\Bigr)^{\ep}\Biggl[
                     \frac{2\beta_{0,Q}}{\ep^2}\gamma_{gq}^{(0)}
                    +\frac{\hat{\gamma}_{gq}^{(1)}}{2\ep}
                    +a_{gq,Q}^{(2)}
                    +\overline{a}_{gq,Q}^{(2)}\ep
                        \Biggr] + O(\varepsilon^2)~. \label{AgqQ2unren1}
  \end{eqnarray}
The constant and $O(\varepsilon)$ contributions  
$a_{gq,Q}^{(2)}$ and $\overline{a}_{gq,Q}^{(2)}$ read
  \begin{eqnarray}
   a_{gq,Q}^{(2)}&=&
          T_FC_F\Biggl\{
            \frac{4}{3}\frac{N^2+N+2}{(N-1)N(N+1)}
               \Bigl(2\zeta_2+S_2+S_1^2\Bigr)
           -\frac{8}{9}\frac{8N^3+13N^2+27N+16}
                            {(N-1)N(N+1)^2}S_1 \nonumber\\ &&
           +\frac{8}{27}\frac{P_1}
                            {(N-1)N(N+1)^3}
                 \Biggr\}~,  \label{AgqQ2van2} \\
\label{agqb}
   \overline{a}_{gq,Q}^{(2)}&=&
          T_FC_F\Biggl\{
            \frac{2}{9}\frac{N^2+N+2}{(N-1)N(N+1)}
               \Bigl(-2S_3-3S_2S_1-S_1^3+4\zeta_3-6\zeta_2S_1\Bigr)\nonumber\\ 
&&
           +\frac{2}{9}\frac{8N^3+13N^2+27N+16}
                            {(N-1)N(N+1)^2}
                \Bigl(2\zeta_2+S_2+S_1^2\Bigr)
           -\frac{4}{27}\frac{P_1S_1}
                             {(N-1)N(N+1)^3}\nonumber\\ &&
           +\frac{4}{81}\frac{P_2}
                             {(N-1)N(N+1)^4}
                \Biggr\}~, \N\\
  \end{eqnarray}
with
  \begin{eqnarray}
   P_1&=&43N^4+105N^3+224N^2+230N+86 \\
   P_2&=&248N^5+863N^4+1927N^3+2582N^2+1820N+496~. 
  \end{eqnarray}
Here $S_{\vec{a}} \equiv S_{\vec{a}}(N)$ denote the (nested) harmonic sums,
\cite{HSUM},
  \begin{eqnarray}
  S_{b,\vec{a}}(N) = \sum_{k=1}^N \frac{({\rm sign}(b))^k}{k^{|b|}} 
S_{\vec{a}}(k)~.
  \end{eqnarray}

\noindent
The renormalized operator matrix element is given by
  \begin{eqnarray}
   A_{gq,Q}&=&
                    a_s^2\Biggl[ \frac{\beta_{0,Q}\gamma_{gq}^{(0)}}{2}
   \ln^2 \Bigl(\frac{m^2_Q}{\mu^2}\Bigr)
   +\frac{\hat{\gamma}_{gq}^{(1)}}{2} \ln \Bigl(\frac{m^2_Q}{\mu^2}\Bigr)
   +a_{gq,Q}^{(2)}-\frac{\beta_{0,Q}\gamma_{gq}^{(0)}}{2}\zeta_2
   \Biggr]+O(a_s^3)~.
   \label{AgqQ2Ren3}
  \end{eqnarray}
Here the anomalous dimensions $\gamma_{gq}^{(0,1)}$ and $\gamma_{gg}^{(0,1)}$
are
\begin{eqnarray} 
\gamma_{gq}^{(0)} &=&-4C_F\frac{N^2+N+2}{(N-1) N (N+1)}~, \\
\gamma_{gg}^{(0)} &=&8C_A\left[S_1-2\frac{N^2+N+1}
                       {(N-1)N(N+1)(N+2)}\right]-2 \beta_0(N_f)~, \\
\hat{\gamma}_{gq}^{(1)}&=&C_FT_F\Biggl(
             -\frac{32}{3}\frac{N^2+N+2}{(N-1)N(N+1)}S_1
             +\frac{32}{9}\frac{8N^3+13N^2+27N+16}{(N-1)N(N+1)^2}
             \Biggr)~, 
\\
\hat{\gamma}_{gg}^{(1)}&=&  8 C_F T_F \left[
                            1
                            - \frac{4}{3} \frac{1}{N-1}
                            + \frac{16}{N}
                            - \frac{6}{N^2}
                            + \frac{4}{N^3}
                            - \frac{8}{N+1}
                            - \frac{10}{(N+1)^2}
                            + \frac{4}{(N+1)^3}
                            - \frac{20}{3} \frac{1}{N+2}\right]
\nonumber\\
&&                          + \frac{16}{3} C_A T_F \left[
                            2
                            + \frac{23}{3} \frac{1}{N-1}
                            - \frac{19}{3} \frac{1}{N}
                            -  \frac{2}{N^2}
                            + \frac{19}{3} \frac{1}{N+1}
                            -  \frac{2}{(N+1)^2}
                            - \frac{23}{3} \frac{1}{N+2}
                            - \frac{10}{3} S_1 \right]~,
\nonumber\\
\end{eqnarray}
and $\hat{\gamma}^{(0)}_{gg} = (8/3) T_F$. A closer look at Eqs. 
(\ref{AgqQ2Ren3},\ref{eqAGG}) reveals, that the terms $\propto \zeta_2$ cancel.
The coefficients of the un-renormalized OME $\hat{A}_{gg,Q}$ are given 
by
  \begin{eqnarray}
   \hat{A}_{gg,Q}^{(1)}&=&
             -\frac{2\beta_{0,Q}}{\ep}f(\ep)~,
                         \label{AggQ1unren1}
\\  
 {\hat{A}}_{gg,Q}^{(2)}&=&
             \Bigl(\frac{m^2_Q}{\mu^2}\Bigr)^{\ep}
                  \Biggl[
                          \frac{1}{2\ep^2}
                             \Bigl\{
                                \gamma_{gq}^{(0)}\hat{\gamma}_{qg}^{(0)}
                               +2\beta_{0,Q}
                                    \Bigl(
                                           \gamma_{gg}^{(0)}
                                          +2\beta_0
                                    \Bigr)
                             \Bigr\}
                         +\frac{\hat{\gamma}_{gg}^{(1)}}{2\ep}
                         +a_{gg,Q}^{(2)}
                         +\overline{a}_{gg,Q}^{(2)}\ep
                 \Biggr]
 \N\\ &&
               +\frac{4\beta_{0,Q}^2f(\ep)^2}{\ep^2}+O(\ep^2)~.
                \label{AggQ2unren1massren}
  \end{eqnarray}
The constant and $O(\varepsilon)$ contributions  
$a_{gg,Q}^{(2)}$ and $\overline{a}_{gg,Q}^{(2)}$ are
  \begin{eqnarray}
\label{eqAGG}
    a_{gg,Q}^{(2)}&=&
      T_FC_A\Biggl\{
                      -\frac{8}{3}\zeta_2S_1
                      +\frac{16(N^2+N+1)\zeta_2}
                            {3(N-1)N(N+1)(N+2)}
                      -4\frac{56N+47}
                             {27(N+1)}S_1 \nonumber\\ &&
                      +\frac{2P_3}
                            {27(N-1)N^3(N+1)^3(N+2)}
         \Biggr\} \nonumber\\ &&
     +T_FC_F\Biggl\{
                      \frac{4(N^2+N+2)^2\zeta_2}
                           {(N-1)N^2(N+1)^2(N+2)}
                      -\frac{P_4}
                            {(N-1)N^4(N+1)^4(N+2)}
         \Biggr\}~, \label{aggQ2}
\end{eqnarray} \begin{eqnarray}
\label{eqAGGB}
   \overline{a}_{gg,Q}^{(2)}&=&
     T_FC_A\Biggl\{
                     -\frac{8}{9}\zeta_3S_1
                     -\frac{20}{9}\zeta_2S_1
                     +\frac{16(N^2+N+1)}
                           {9(N-1)N(N+1)(N+2)}\zeta_3 
                     +\frac{2N+1}
                           {3(N+1)}S_2
                     -\frac{S_1^2}{3(N+1)} \nonumber\\ &&
                     +\frac{4P_5\zeta_2}
                           {9(N-1)N^2(N+1)^2(N+2)}
                     -2\frac{328N^4+256N^3-247N^2-175N+54}
                            {81(N-1)N(N+1)^2}S_1 
\nonumber\\ 
&&
                     +\frac{P_6}
                           {81(N-1)N^4(N+1)^4(N+2)}
         \Biggr\} \nonumber
\end{eqnarray} \begin{eqnarray}
&&    +T_FC_F\Biggl\{
                       \frac{4(N^2+N+2)^2\zeta_3}
                            {3(N-1)N^2(N+1)^2(N+2)}
                      +\frac{P_7\zeta_2}
                            {(N-1)N^3(N+1)^3(N+2)} \nonumber\\ &&
                      +\frac{P_8}
                             {4(N-1)N^5(N+1)^5(N+2)}
         \Biggr\}~, \label{abarggQ2}
  \end{eqnarray}
where
  \begin{eqnarray}
   P_3&=&15N^8+60N^7+572N^6+1470N^5+2135N^4+1794N^3+722N^2-24N-72~,\\
   P_4&=&15N^{10}+75N^9+112N^8+14N^7-61N^6+107N^5+170N^4
   +36N^3 \nonumber\\ &&
   -36N^2-32N-16~,\\ 
   P_5&=&3N^6+9N^5+22N^4+29N^3+41N^2+28N+6~,\\
   P_6&=&3N^{10}+15N^9+3316N^8+12778N^7+22951N^6+23815N^5+14212N^4+3556N^3
      \N\\ && -30N^2+288N+216~,\\
   P_7&=&N^8+4N^7+8N^6+6N^5-3N^4-22N^3-10N^2-8N-8~,\\
   P_8&=&31N^{12}+186N^{11}+435N^{10}+438N^9-123N^8-1170N^7-1527N^6-654N^5
      \N\\ && +88N^4 -136N^2-96N-32~.
  \end{eqnarray}
The renormalized operator matrix element $A_{gg,Q}$ reads
  \begin{eqnarray}
   A_{gg,Q} &=& a_s \frac{4}{3}T_F\ln \Bigl(\frac{m^2_Q}{\mu^2}\Bigr) 
   + a_s^2 \Biggl[
      \frac{1}{8}\Biggl\{
                          2\beta_{0,Q}
                             \Bigl(
                                    \gamma_{gg}^{(0)}
                                   +2\beta_0 
                             \Bigr)
                         +\gamma_{gq}^{(0)}\hat{\gamma}_{qg}^{(0)}
                 \Biggr\}
                    \ln^2 \Bigl(\frac{m^2_Q}{\mu^2}\Bigr)
                +\frac{\hat{\gamma}_{gg}^{(1)}}{2}
                     \ln \Bigl(\frac{m^2_Q}{\mu^2}\Bigr)
\N\\ &&
                +a_{gg,Q}^{(2)}
                -\frac{\zeta_2}{8}\Bigl[
                                         2\beta_{0,Q}
                                            \Bigl(
                                                   \gamma_{gg}^{(0)}
                                                  +2\beta_0
                                            \Bigr)
                                        +\gamma_{gq}^{(0)}
                                            \hat{\gamma}_{qg}^{(0)}
                                  \Bigr]
                \Biggr]+O(a_s^3)~.   \label{AggQ2Ren2}
  \end{eqnarray}
We agree with the results for $a_{gq,Q}^{(2)}$ and $a_{gg,Q}^{(2)}$ 
given in \cite{BUZA2}, which we presented in (\ref{AgqQ2van2},\ref{aggQ2}).
The new terms
$\overline{a}_{gq,Q}^{(2)}$ and $\overline{a}_{gg,Q}^{(2)}$, (\ref{agqb}, 
\ref{abarggQ2}), contribute to all OMEs $A_{ij}^{(3)}$ through 
renormalization.
With respect to the mathematical structure, $a_{gq(gg),Q}^{(2)}$ and 
$\overline{a}_{gq(gg),Q}^{(2)}$,
(\ref{AgqQ2van2},\ref{aggQ2},\ref{agqb},\ref{abarggQ2}), belong to the class
being observed for two--loop corrections before, \cite{JB08}. In the present 
case even only single harmonic sums contribute.
We checked our results for the moments $N=2, \ldots, 8$ using the code 
{\tt MATAD},~\cite{MATAD}. An additional check is provided by the sum 
rules in Eqs. (\ref{sumrule1},\ref{sumrule2}), which are fulfilled by the 
renormalized OMEs presented here and in Refs. \cite{BUZA1,BUZA2,BBK1}. 
Moreover, we observe that these rules are obeyed on the unrenormalized level
as well, even up to $O(\ep)$, \cite{BBK2}.

To describe the evolution of the parton distributions, 
Eqs.~(\ref{HPDF1}--\ref{HPDF2}), the OMEs (\ref{AgqQ2Ren3},\ref{AggQ2Ren2})
have to be  supplemented by the corresponding 1PR terms
\begin{eqnarray}
\label{AA1}
A_{Qg}^{(2)} &\rightarrow& A_{Qg}^{(2)} + 
a_s^2 4 \beta_{0,Q} T_F \frac{N^2+N+2}{N(N+1)(N+2)} 
\ln^2\left(\frac{\mu^2}{m_Q^2}\right)~,\\
\label{AA2}
A_{gg,Q}^{(2)} &\rightarrow& A_{gg,Q}^{(2)} + 
a_s^2  \beta_{0,Q}^2 \ln^2\left(\frac{\mu^2}{m_Q^2}\right)~.
\end{eqnarray}
Eqs.~(\ref{AA1},\ref{AA2}) agree with the results presented in 
Ref.~\cite{BUZA2}.
In applying these parton densities in other hard scattering processes this 
modification also affects part of the massless hard scattering cross sections 
there, as outlined above.

\section{Conclusions}
\label{sec-4}

\vspace{1mm}\noindent
We calculated the massive gluonic operator matrix elements $A_{gq,Q}$ and 
$A_{gg,Q}$, being required in the description of heavy flavor parton densities
at scales sufficiently above threshold, to $O(a_s^2 \varepsilon)$. We confirm
previous results given in \cite{BUZA2} for the constant terms and obtained 
newly the $O(\varepsilon)$ terms which enter the 3-loop corrections to 
$A_{ij}$ via renormalization. We reminded of details of the charge 
renormalization and clarified that  additional terms at $O(a_s^2)$ are to be 
included in the data analysis in the FFNS and VFNS using the $\overline{\rm MS}$ 
scheme. 

\vspace{5mm}\noindent
{\bf Acknowledgments.}~~We would like to thank S. Alekhin and E. Laenen for  
useful discussions. This work was supported in part by DFG 
Sonderforschungsbereich 
Transregio 9, Computergest\"utzte Theoretische Teilchenphysik, Studienstiftung 
des Deutschen Volkes, the European Commission MRTN HEPTOOLS under 
Contract No. MRTN-CT-2006-035505, the Ministerio de Ciencia e Innovacion 
under Grant No. FPA2007-60323, CPAN (Grant No. CSD2007-00042), the 
Generalitat Valenciana under Grant No. PROMETEO/2008/069, and by the European 
Commission MRTN FLAVIAnet under Contract No. MRTN-CT-2006-035482.

\newpage

\end{document}